# Terahertz non-label subwavelength imaging with composite photonics-plasmonics structured illumination

*Jin Zhao, Li-Zheng Yin, Feng-Yuan Han, Yi-Dong Wang, Tie-Jun Huang, Chao-Hai Du, and Pu-Kun Liu\**

State Key Laboratory of Advanced Optical Communication Systems and Networks, Department of Electronics, Peking University, Beijing 100871, China

*Corresponding author: pkliu@pku.edu.cn

**Abstract**

Inspired by the capability of structured illumination microscopy (SIM) in subwavelength imaging, many researchers devoted themselves to investigating this methodology. However, due to the free-propagating feature of the traditional structured illumination fields, the resolution can be only improved up to double times compared with the diffraction-limited microscopy. Besides, most of the previous studies, relying on incoherent illumination sources, are restricted to fluorescent samples. In this work, a subwavelength non-fluorescent imaging method is proposed based on the terahertz traveling wave and plasmonics illumination. Excited along with a metal grating, the spoof surface plasmons (SSPs) are employed as the plasmonics illumination. When the scattering waves with the SSPs illumination are captured, the sample's high-order spatial frequencies (SF) components are already encoded into the obtainable low-order ones. Then, a post-processing algorithm is summarized to shift the modulated SF components to their actual positions in the Fourier domain. In this manner, high-order SF components carrying the fine information are introduced to reconstruct the desired imaging, leading to an improvement of the resolution up to $0.12\lambda_0$. Encouragingly, the resolution can be further enhanced by tuning the working frequency of the SSPs. This method holds promise for some important applications in terahertz non-fluorescent microscopy and sample detection with weak scattering.

**Introduction**

Optical microscopy has become an essential tool in a wide range of applications, including biological imaging, material identification, device creation, etc. However, in conventional far-field microscopy, one of the major limitations is the insufficient resolution, which is constrained to approximately half the working wavelength of the illumination light [1]. This phenomenon is termed the Rayleigh limit or classical diffraction limit [2]. Specifically, the samples' fine information is encompassed in the high-order spatial frequency (SF) components, manifesting as the evanescent waves. They reduce exponentially perpendicular to the propagating direction [3]. Thus, the high-order SF components permanently vanish in the far-field, posing a wavelength level resolution in the ideal circumstance [4].

To overcome this resolution limitation, a series of optical strategies have been developed over the past decades. For instance, the super-resolution fluorescence microscopy techniques, including stimulated emission depletion [5,6], stochastic optical reconstruction [7,8], structured illumination microscopies (SIM) [9-11], have endowed a breakthrough in subwavelength resolution imaging. Among the myriad fluorescence super-resolution approaches, SIM is considerably remarkable due to its fast and highly parallelizable feature [12]. In SIM, a finely structured illumination pattern is applied to illuminate the sample. In this case, the high-order SF components (corresponding to the fine detail of the sample) can be down-modulated into the passband of the objective and turn out to be detectable. By post-processing, the high-order SF information can be precisely shifted and located at the true position in the spatial frequency domain ($k$-domain). As a result, one can defeat the diffraction barrier and finally achieve a subwavelength resolution.

However, owing to the free-propagating feature of the structured illumination fields, the spatial resolution of traditional SIM is restricted to a two-fold enhancement in comparison with the diffraction-limited techniques [13]. Aside from the resolution, the application scenarios of SIM should be taken into consideration. Most of the aforementioned researches depend on incoherent imaging techniques and fluorescent samples, which become out-of-operation for label-free imaging [14-19]. From the perspective of the terahertz range, the situation is similar. The existing researches focus on the fluorescent samples and obtain an excellent resolution for subwavelength imaging [20]. However, to the best of our knowledge, the terahertz label-free subwavelength imaging with structured illumination method remains incomplete.

In this work, a terahertz subwavelength imaging method of label-free samples is proposed based on composite photonics-plasmonics structured illumination. The spoof surface plasmons (SSPs) subsisted on a metal grating are employed for the plasmonics illumination. When the sample is impinged by the SSPs, the scattering field is encoded. That is, the high-order SF components of the sample (carrying fine information) are down-transformed to the low-order and obtainable SF ones. Then, the scattering field is caught to get the modulated high-order SF information. Similarly, with vertical and oblique incidences of the traveling waves (photonics illuminations), low-order SF information of the sample is also acquired. Subsequently, an iterative algorithm is applied to extract and locate all the SF components in their true positions. In this way, a wide SF spectrum is generated, which is imperative for subwavelength imaging. Based on this principle, the simulated resolution of the proposed method reaches up to $0.12\lambda$. Note that the resolution can be further improved by elaborately design the SSPs' working frequency. This method is beneficial for applications of terahertz nonfluorescent microscopy and weak scattering sample detection.

**Theory**
**1.** *Imaging Principle*

The imaging principle in this work relies on the coherent scattering from the label-free samples. Before elaborating on the imaging principle, there are several settings to be clarified. Firstly, this process is between the Fourier domain ($k$-domain) and the spatial domain ($x$-$y$), where $k$ is the SF vector. Here, we consider the condition where the sample is positioned along the $x$-direction and infinite in the $y$-direction. The incident wave is on the $x$-$z$ plane, lacking the component $k_y$. Secondly, we postulate that when an incident wave impinges on a thin subwavelength sample, the scattered wave can be illustrated as $s(x) = o(x)\exp(ik_t x)$. $o(x)$ refers to the sample's complex field [21] and $k_t$ is the SF component of the incident wave along the $x$-axis. Thirdly, the coherent point spread function (PSF) $h_c(x)$ of our imaging system is defined via its Fourier transform $H_c(k_x)$. $H_c(k_x)$ is modeled by an abrupt rectangular low-pass filter, which has a width of $2k_c$ ($k_c$ is the cut-off frequency). Since the imaging system of this work is in the free space, $k_c$ can be substituted by $k_0$ (the wavenumber of vacuum). $H_c(k_x)$ is rendered into gray in Fig.1 (a), where $C$ denotes the module value of the SF component. On this basis, the detected intensity imaging of the sample is indicated as $i(x)$ and derived as the following equation[21]

$$i(x) = |h_c(x) \otimes s(x)|^2, \qquad (1)$$

where $\otimes$ is the convolution operator. Eq. (1) can be written in the $k$-domain, which is signified with Eq. (2)

$$I(k_x) = autocorr[H_c(k_x)S(k_x)], \qquad (2)$$

where $I(k_x)$ and $S(k_x)$ are the SF distributions of the detected image intensity, and the scattering field, respectively. It should be noticed that $S(k_x) = O(k_x - k_t)$, where $O(k_x)$ is the SF distribution of the sample's field. Eq. (2) thus can be derived as

$$I(k_x) = autocorr[H_c(k_x)O(k_x - k_t)]. \qquad (3)$$

In this manner, the spatial spectrum of the sample is shifted to $k_x'$, manifesting as $k_x' = k_x - k_t$. The detected image intensity $I(k_x)$ contains the attainable SF information, which is governed by $|k_x'| < k_c$. Since $k_c = k_0$, it is written as $-k_0 + k_t < k_x < k_0 + k_t$.

According to the aforementioned principle, the conversion of the collected SF range is attributed to the tangential component $k_t$ of the illuminating wave. When a plane wave impinges the sample vertically, the obtained SF region is $-k_0 < k_x < k_0$ (same as the system's passband), as shown in Fig. 1(a). Analogously, a plane wave with left oblique illumination leads to the acquisition of the SF spectrum in $-k_0 - k_0\sin\theta < k_x < k_0 - k_0\sin\theta$ ($\theta$ is the incident angle). For the right oblique incidence, the acquired range is $-k_0 + k_0\sin\theta < k_x < k_0 + k_0\sin\theta$ (Fig. 1(b)). Since $\theta < 90°$, the detectable region is limited by $\pm 2k_0$.

To further increase the observable regime, the evanescent wave is exploited as the illuminated wave (along both $x$ and $-x$-directions). The tangential component $k_{eva}$ of the evanescent wave is larger than $k_0$. Consequently, the attainable SF regime ($-k_0 \pm k_{eva} < k_x < k_0 \pm k_{eva}$) is extended further, as depicted in Fig. 1(c).

Finally, through judiciously tailoring the incident angle of the plane wave and the tangential component of the evanescent wave, a wide integrated $k$-domain information can be spliced with no empty regions. It is noted that the wide information in the $k$-domain is imperative for high-resolution imaging. Thus, applying the inverse Fourier transform, the subwavelength spatial image of the sample is restored.

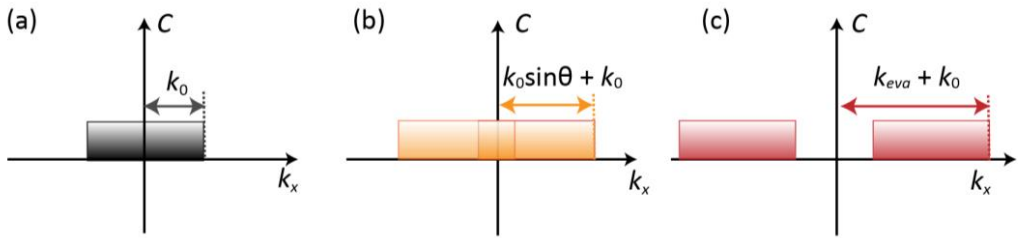

**Figure 1.** Detectable SF ranges under different illuminated patterns. (a) Vertical illumination. (b) Oblique illuminations. (c) Evanescent illuminations.

## 2. *Reconstrcution algorithm*

After gathering all of the raw intensity images under different illuminated patterns, a meliorated Gerchberg-Saxton algorithm [22,23] is exploited to iteratively generate a wide $k$ spectrum. In this way, a subwavelength-resolution image can be realized. The procedure involves five steps.

*Step* 1. An initial value $I_s^{1/2}e^{i\varphi}$ of the subwavelength sample in the spatial domain is assumed, using $\varphi = 0$ and $I_s$ being a constant value (detected data group is also available [1]). Applying the Fourier transform for the initial value, a wide band spectrum $W(k_x)$ can be presented in the $k$-domain

$$W(k_x) = \mathscr{F}\left(\sqrt{I_s}e^{i\varphi}\right). \tag{4}$$

*Step* 2. The low SF components are restored with the image under vertical illumination, i.e., oblique illumination with an incident angle of 0°. A subregion of the wide $k$-domain is extracted and dealt with the invert Fourier transform, producing a new temporary image $I_t^{1/2}e^{i\varphi_t}$ The selected area in the $k$-domain hinges on the low-pass filter $H_c(k_x)$. The location of the low-pass filter depends on the tangential wavenumber of the illumination pattern. Under the vertical incidence (tangential wavenumber being zeros), the position of the filter's center is $k_x = 0$. Thus, the domain of this subregion is $-k_0 < k_x < k_0$. Resultantly, the aforementioned procedure is explicated as

$$\sqrt{I_t}e^{i\varphi_t} = \mathscr{F}^{-1}\left[W(k_x)H_c(k_x)\right], \tag{5}$$

where $\mathscr{F}^{-1}$ indicates the inverse Fourier transform. Then, the amplitude $I_t^{1/2}$ of the temporary image $I_t^{1/2}e^{i\varphi_t}$ is substituted by the detected image amplitude $I_d^{1/2}$. Next, we use the Fourier transform of $I_d^{1/2}e^{i\varphi_t}$ to replace the corresponding subregion in the $k$-domain. An updated SF spectrum $W_{new}(k_x)$ is configured

$$W_{new}(k_x) = W(k_x)\left[1 - H_c(k_x)\right] + H_c(k_x)\mathscr{F}\left(\sqrt{I_d}e^{i\varphi_t}\right). \tag{6}$$

*Step* 3. By repeating step 2 (select a small, circular region of $k$-domain and update it by the corresponding detected data), the high SF components are also obtained with oblique and evanescent illuminations. For the left and right oblique illumination with incidence angle $\theta$, the extracted subregion of $k$-domain is displayed as $-k_0 - k_0\sin\theta < k_x < k_0 - k_0\sin\theta$ and $-k_0 + k_0\sin\theta < k_x < k_0 + k_0\sin\theta$, respectively. Similarly, the extracted subdomain with the evanescent illumination is presented as $-k_0 \pm k_{eva} < k_x < k_0 \pm k_{eva}$, where $k_{eva}$ is the evanescent wave's component along the $x$-direction.

*Step* 4. Repeating steps 2 and 3 until a self-consistent solution is realized.

*Step* 5. The converged solution $W_{solved}(k_x)$ is processed with an inverse Fourier transformation to recover a subwavelength image $i_{solved}(x)$. After an origin calibration, a subwavelength-resolution image is produced. The complete pseudocode is demonstrated in the supplementary material.

## 3. *SSPs Illumination Field*

As mentioned above, evanescent waves are applied as the illumination field to achieve a down-modulation of the high SF information. In the optical region, the surface plasmon polaritons (SPPs), produced by collective oscillations of electron gases near the metallic surfaces and interfaces [24], can be utilized as evanescent wave illumination [1,25-26]. However, metals display the ideal electric conductor property in the terahertz zone, as a result of which natural SPPs are severely declined [27]. To solve this problem, the evanescent wave illumination is achieved by the SSPs in this work. The SSPs propagate on a metal grating surface and decay exponentially along the vertical direction [28,29]. Via manipulating the working frequencies of the SSPs, a wealth range of tangential wavenumbers ($k_{eva}$) are available, benefiting the system design.

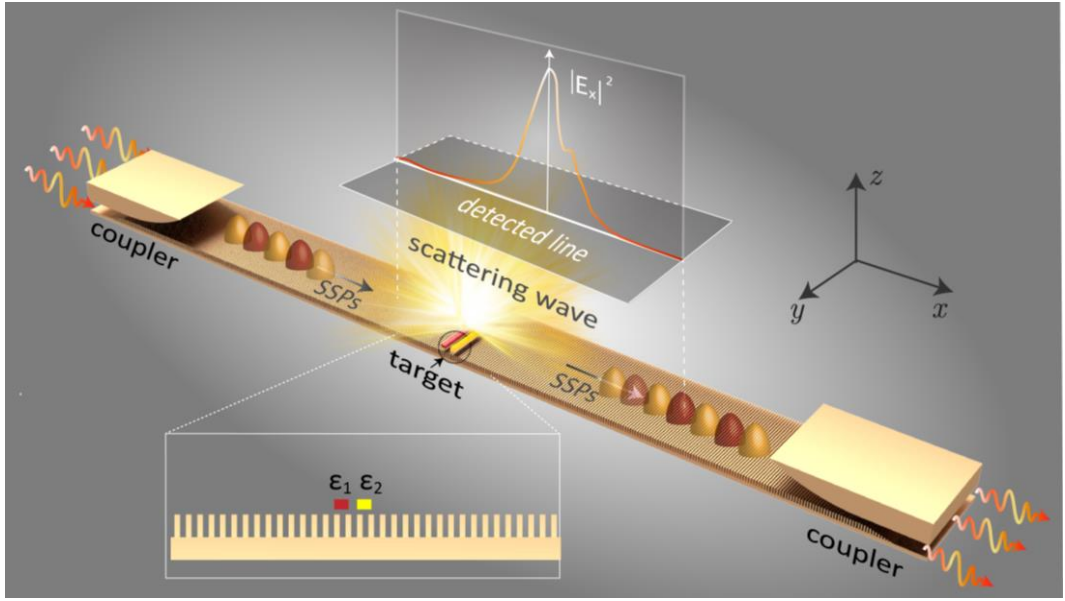

**Figure 2.** Schematic diagram of the SSPs illumination, where $E_x$ is the $x$-direction component of the scattering field.

The schematic diagram of the proposed approach is shown in Fig. 2, wherein a metal grating is applied to sustain the SSPs on the surface. Two thin samples with different relative permittivities are deposited above the grating in a subwavelength distance. Leveraging on a coupler between the SSPs and traveling waves, the SSPs are excited adequately in this device. Then, the SSPs propagate along the grating and impinge on the samples, rendering a detectably scattering wave. After that, the SSPs propagate forward and exit as a traveling wave through another coupler. It is worth emphasizing that the SSPs can transmit along both the $x$-direction and $-x$-direction, just depending on which coupler it is excited on.

Specifically, the grating's unit cell is presented in Fig. 3(a), where $h$ = 0.049 mm, $p$ = 0.02 mm, and $d$ = 0.01 mm (the same size in all simulations). The dispersion of the SSPs contains the fundamental mode and the high-order modes [30]. To prevent interference from multiple modes, only the fundamental mode [Fig. 3(a)] is employed in this work. In this work, the working frequencies of SSPs are $f_1$ = 1.1 THz, $f_2$ = 1.23 THz and $f_3$ = 1.33 THz. Corresponding to the working frequencies, the tangential wavenumbers (x-direction) are $k_{eva1}$ = 1.387$k_0$, $k_{eva2}$ = 2.082$k_0$, $k_{eva3}$ = 3.539$k_0$. $k_0$ is caculated as $2\pi f_0/c$, where $c$ is the velocity of light. $f_0$ is the working frequency of the vertical and oblique plane waves, set as 1.185 THz. As the working frequencies increasing, $k_{eva}$ becomes larger, resulting in a more confined SSP field [Figs. 3(b), (c), and (d)]. Notably, further promotion of the imaging resolution is feasible by adding new SSPs illumination with high working frequency. With a higher working frequency, larger SF components can be down-modulated into the passband of the system. Thus, it leads to a larger effective numerical aperture NA = (max{$k_{eva1}$, $k_{eva2}$, $k_{eva3}$, $k_{eva4}$, … } + max{$f_{eva1}$, $f_{eva2}$, $f_{eva3}$, $f_{eva4}$, … } / $f_0$)$k_0$ and a higher resolution [calculated as $\lambda_0/(2NA)$]. Since more illumination patterns causes a more sophisticated post-process, a balance between the resolution and number of illumination patterns should be elaborately determined.

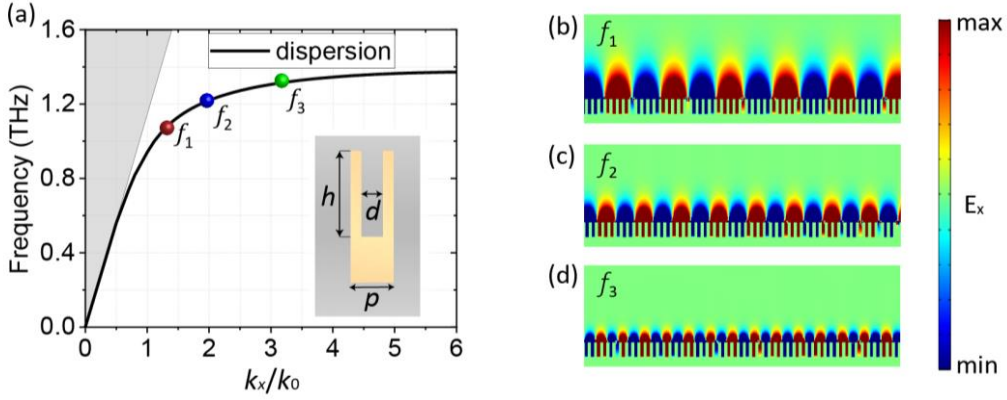

**Figure 3.** Dispersion of the grating and the SSPs field of different working frequencies. (a) Unit structure and fundamental mode dispersion curve of the unit. (b) SSPs field at $f_1$. (c) SSPs field at $f_2$. (d) SSPs field at $f_3$.

**Simulation and Results**

Hereto, the procedure of how to recover a subwavelength image in the terahertz region is perspicuous. To encode the Fourier spectrum, there are various illumination patterns, including vertical, oblique, and evanescent illuminations. For the data collection, the scattering field intensity of the sample is directly detected, which avoids extra laser to excite the sample [20]. For the image reconstruction, a post-process algorithm is applied.

With the assistance of the commercial software COMSOL Multiphysics 5.4a, the scattering fields intensity are investigated under diversiform illuminations. Take the single sample (0.15$\lambda_0$×0.1$\lambda_0$ on the x-z plane) imaging as an example, for the vertical illumination, the sample is directly impinged by a plane wave [see Fig. 4(a)]. Similarly, plotted by Figs. 4(b) and (c), the oblique patterns employ plane waves with ±66° incident angles. The corresponding tangential wavenumber are ±0.914 $k_0$. For the evanescent illumination, the sample is placed above the grating surface with an extremely close distance, which is approximately 0.024$\lambda_0$–0.12$\lambda_0$. Figs. 4(d) and (e) demonstrate the scattering fields intensity with SSPs propagating along x and –x-directions (working frequency is $f_1$). In addition, the SSPs illuminations at $f_2$ and $f_3$ are also simulated. Consequently, the tangential wavenumbers of the nine illumination waves are [0, 0.914$k_0$, −0.914$k_0$, 1.387$k_0$, −1.387$k_0$, 2.082$k_0$, −2.082$k_0$, 3.539$k_0$,

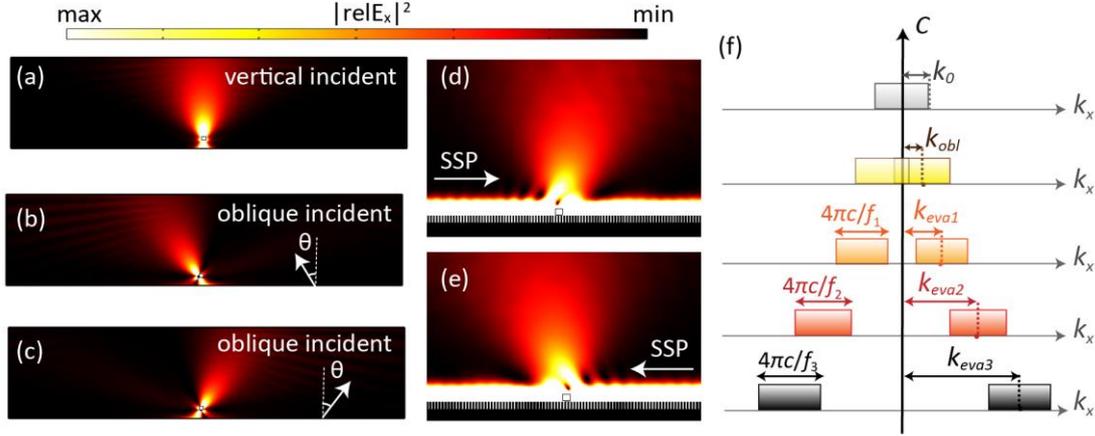

**Figure 4.** Scattering fields intensity with various illuminations for the corresponding SF regions. (a) Scattering field with the vertical illumination. (b) Right oblique illumination. , (c) Left oblique illumination. (d) SSPs illumination along with the *x*-direction (working frequency is $f_1$). (e) SSPs illumination along with the −*x*-direction (working frequency is $f_1$). (f) Corresponding detectable SF regimes of diverse illuminated patterns.

−3.539$k_0$]. They represent centers of the detectable SF regions under respective illumination patterns. In Fig. 4(f), the nine obtained SF spectrums are presented. Since the working frequencies of propagating waves and SSPs are distinct, the corresponding half-width of the detectable SF regions are derived as ($f_i$ / $f_0$) $\times$ $k_0$ = 2$\pi c$ / $f_i$, $i$ = 0, 1, 2, and 3.

Next, the scattering fields intensity under these nine illumination patterns are collected, without the procedure of laser excitation. For the resolution investigation, the length of the detective line on the *x-z* plane is 2540 μm, which is 268 μm away from the sample. With the post-process algorithm, a wide *k* spectrum is reconstructed. Then the *k* spectrum is processed by the apodization and zero-padding function to extract the critical SF information and reveal smooth results, respectively.

For the recovered results, the first demonstration is the imaging of a single, isolated sample with the subwavelength feature, whose dimension is 0.15$\lambda_0$ × 0.1$\lambda_0$ on the *x-z* section. The relative permittivity of the sample is $\varepsilon$ = 2.5, characterizing the dielectric property. The reconstructed images of this sample are depicted in Fig. 5(a), which contain various illumination groups. The black curve illustrates the result generated by the whole nine illuminations (group1), presenting a 0.11$\lambda_0$ full width at half maximum (FWHM). When the SSPs illuminations at $f_2$ and $f_3$ are eliminated in the reconstruction process (group2), the FWHM becomes larger than before, as designated by the blue curve. Ulteriorly, if only the vertical and oblique illuminations (group3) remain without any SSPs illumination, the FWHM increases further. The modulus of the *k* spectrum is plotted in Fig. 5(b), corresponding to these assorted illumination combinations. The orange region represents the attainable SF spectrum with vertical and oblique patterns. The blue plus orange regions are the whole achieved SF information under the group2 illumination. When the combination is group1, the homologous SF region includes the blue, orange, and gray pieces. Apparently, high SF information in the $k$-domain is removed without the relevant SSPs illumination, leading to a wide FWHM. From the point of Abbe's criterion (imaging a single sample) [31], this phenomenon is rational. According to the criterion, smaller FWHM means higher resolution, which is attained through collecting more fine information (high SF regions). Since the SSPs illumination brings high SF in $k$-domain, abandoning them naturally causes the low resolution and large FWHM. Finally, based on Abbe's criterion, this work can realize a $0.11\lambda_0$ resolution, which can be further improved via adding SSPs illumination of large $k_{eva}$.

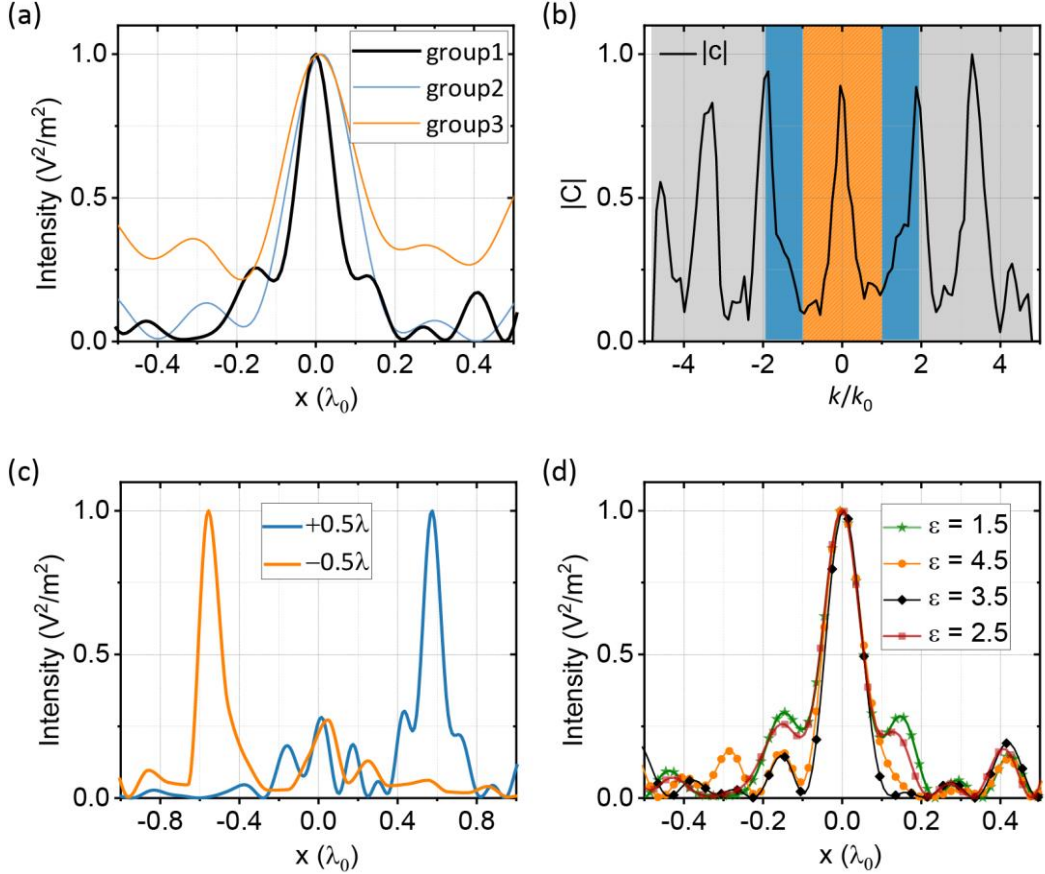

**Figure 5.** Simulated results of an isolated sample. (a) images with three illuminated wave groups, group1, 2, and 3 respectively corresponding to the only vertical and oblique, all patterns without SSPs illuminations at $f_2$ and $f_3$, and all the illuminations. (b) modulus value of the Fourier spectrum. (c) images with ±0.5λ displacement. (d) images with different relative permittivities.

Besides, to verify the robustness of this method, the sample is repositioned along the $x$-axis. Assuming the sample is placed at the origin for the aforementioned discussion. Whereupon we move it to $\pm 0.5\lambda_0$ positions, keeping the other conditions fixed. The result is displayed in Fig. 5(c), which testifies the method's stability with disparate locations. There exists a slight difference between the two curves, which is mainly caused by the personalized iteration number when a convergent solution is generated. Another issue needed to be explored is whether the sample's dielectric feature has an impact on the reconstruction. The relative permittivity is set as 1.5, 2.5, 3.5, and 4.5 with the other conditions being identical. Comfortingly, Fig. 5(d) illustrates that the restored imaging keeps stable, meaning that the method is robust to the variation of the relative permittivity within a certain range.

Up to here, we have demonstrated the $0.11\lambda_0$ resolution and the robustness of this method. However, there are many other criteria to evaluate the imaging resolution in the literature. The most likely accepted is the Rayleigh criterion [2], which confirms the minimum resolvable distance by the imaging of two closely separated samples. In the following, the examination of the resolution in the view of the Rayleigh criterion is presented.

First, we consider two same samples with $\varepsilon = 2.5$, $h = 0.1\lambda_0$, and width $w_s$. They are separated by a gap, whose length is $w_g = w_s$. We recover the images with $w_s$ ranging from $0.15\lambda_0$ to $0.225\lambda_0$. Figs. 6(a)-(c) depict the images under these conditions, which confirm the method's subwavelength imaging ability of two samples. Furthermore, in Fig. 6(d), the samples are set extremely small with $h = w_s = 0.02\lambda_0$ for the judgment of the highest resolution. The distance between the center of them is $w_c = 0.12\lambda_0$. In this case, the samples can be approximately regarded as two points with a separate distance approximately being $0.12\lambda_0$. To increase the interaction between the tiny samples and the illuminating waves, their relative permittivities are raised to 12. Under this condition, the result verifies that a resolution of $0.12\lambda_0$ can be achieved. Subsequently, when the samples get closer to each other ($0.11\lambda_0$), the image becomes inexplicit. As plotted in Fig. 6(d), one peak is almost a quarter lower than the other, illustrating the method cannot distinguish two samples separated by $0.11\lambda_0$. Thus, on basis of the Rayleigh criterion, the ultimate resolution of this method is $0.12\lambda_0$. Actually, there exists a slight deviation about the resolution with the Abbe and Rayleigh criterion. One reason is that the samples in the Rayleigh criterion has a $0.02\lambda_0$ width, and the gap between them is actually shorter than $0.12\lambda_0$. Besides, the undesired reflection between the two samples convey noise to the reconstruction and lead to a slight deviation.

Hereto, the theoretical resolution and the simulated resolution can be exhibited and compared. For this work, the effective NA is decided by the maximum $k_{eva}$ of the SSPs illumination, which is $3.539k_0$. Thus, the highest detectable SF in $k$-domain is $\pm(3.539 + f_3/f_0) k_0 = \pm4.66k_0$, i.e., NA = $4.66k_0$. In this way, the theoretical resolution is approximately $0.107\lambda_0$. According to Abbe's criterion, the simualted resolution is $0.11\lambda_0$. Considering the approximation error, it is almost consistent with the theoretical value. With the Rayleigh criterion, the established resolution is $0.12\lambda_0$, which has a slight excursion. The offset may caused by the finite size of the samples and the undesired reflection between them. Finally, the imaging resolution is demonstrated as $0.12\lambda_0$, almost four times that of diffraction-limited imaging.

Besides, we also restore images of two samples with various relative permittivities. At the beginning, we employ two samples ($h = 0.1\lambda_0$, $w_s = 0.15\lambda_0$, $w_g = 0.2\lambda_0$) with the same relative permittivity $\varepsilon = 2.5$ as a reference standard, presented in Fig. 7(a). Afterward, the relative permittivities of the left and right samples are stated as $\varepsilon_1$ and $\varepsilon_2$, respectively. In Fig. 7(b), it is displayed that when $\varepsilon_1 = 2.5$ and $\varepsilon_2 = 4.5$, the right peak's height is larger than the left one. Analogously, for $\varepsilon_1 = 2.5$ and $\varepsilon_2 = 1.5$, the left peak is higher than the right peak, shown in Fig. 7(c). Actually, due to the sample's different scattering power with diverse relative permittivities, there emerges unequal peaks. Intuitively, a large relative permittivity yields strong scattering under the impinging wave. To some extent, this method can distinguish two samples with distinct relative permittivities, which may be helpful for the subwavelength dielectric determination. It should be mentioned that the ambiguous image in Fig. 6(d) also has unequal peaks, demonstrating the resolution limit. This means when one peak behaves lower than the other, the reason is not only the inconsistent $\varepsilon$. Thus, this function of relative permittivity distinction can be exploited to two same-size samples with relatively large interval.

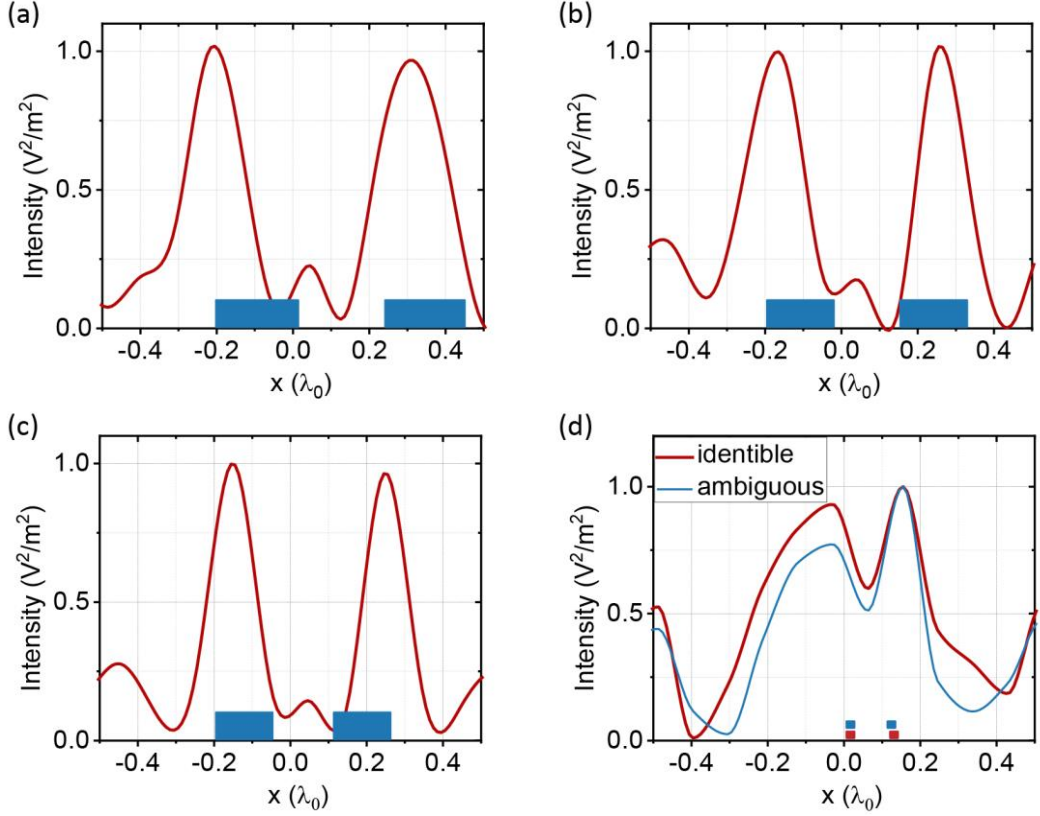

**Figure 6.** Verification of two samples' imaging capability. (a) $w_g = w_s = 0.225\lambda_0$. (b) $w_g = w_s = 0.175\lambda_0$. (c) $w_g = w_s = 0.15\lambda_0$. (d) $w_s = 0.02\lambda_0$, $w_g = 0.12\lambda_0$ (red curve and samples), and $w_g = 0.11\lambda_0$ (blue curve and samples).

Finally, the imaging ability of three samples is also demonstrated in Fig. 7(d). Under this condition, the detected line is moderately adjusted and extended to fit the extended space of the samples. There are three same samples ($h = 0.1\lambda_0$, $w_s = 0.15\lambda_0$, $\varepsilon = 2.5$) with the gap distances $w_{s1} = 0.25\lambda_0$ and $w_{s2} = 0.45\lambda_0$. Consequently, they are successfully reconstructed in the recovery image.

**Discussion**
Firstly, the resolution can be further improved via adding new SSPs illumination of high frequency. In this manner, the effective NA is increased so that the resolution can be enhanced. Remarkably, the working frequencies need to be advisably chosen to ensure a wide *k* spectrum without interrupted region [22]. Secondly, the sample's scattering field intensity is collected straightly, circumventing an extra laser excitation process. Thirdly, the noise in this system may cause side lobes and artifacts in the reconstructed imaging, which can be further reduced through an optimization of the algorithm [32]. Besides, the influence of couplers' reflection and the higher harmonics on the grating can also be sufficiently suppressed to improve the imaging performance. Fourthly, the method can be exploited to the two-dimensional imaging, the key step of which is to realize multidirectional SSPs illuminations on the *x-y* plane. One can rotate the grating around the *z*-axis and simultaneously fix the sample to acquire the multidirectional SSPs. In addition, the multidirectional SSPs illuminations may also be achievable based on an integrated multi-port waveguide [33,34].

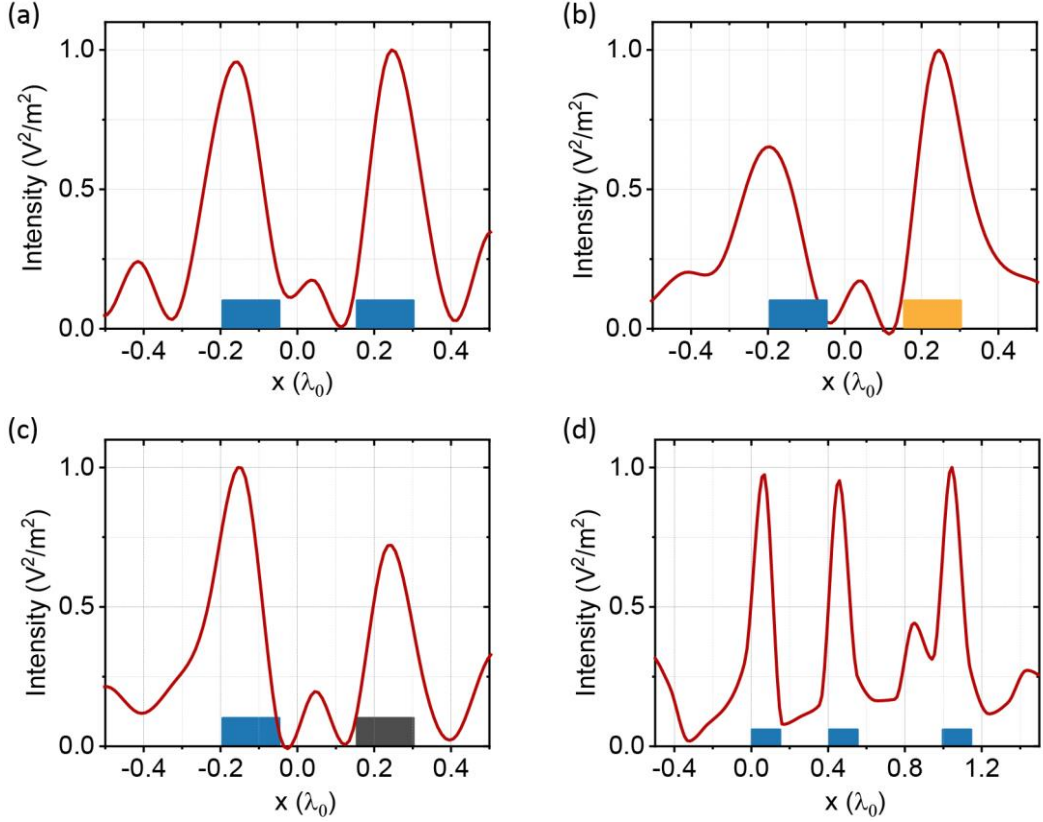

**Figure 7.** Discussion on imaging properties. (a) Image of two identical samples. (b) Image of two samples with $\varepsilon_1 = 2.5$ and $\varepsilon_2 = 4.5$. (c) Image of two samples with $\varepsilon_1 = 2.5$ and $\varepsilon_2 = 1.5$. (d) Image of three same samples with $w_{s1} = 0.25\lambda_0$ and $w_{s2} = 0.45\lambda_0$.

**Conclusion**
In conclusion, this work presents a theoretical proposal to realize a subwavelength coherent imaging method for non-label samples in the terahertz region. A tunable composite photonics-plasmonics illumination pattern is used in the scheme. Depending on the SSPs on a metal grating, the plasmonics illumination pattern is obtained. When the SSPs impinge on the sample,the high-order SF information is down-modulated and contained in the detected scattering field intensity. Accordingly, under the vertical and oblique propagating waves (photonics) illumination, the relatively low-order SF information is also captured. In this manner, an integral wide Fourier spectrum of the sample is achieved, then a subwavelength recovery image is achieved by a post-process algorithm. With nine illumination patterns, a resolution up to $0.12\lambda_0$ is verified against both Abbe's and Rayleigh's criteria. Besides, the robustness of this proof-of-concept method for different samples' parameters is analyzed. Finally, improvements on the resolution and image quality are also discussed for further work. We believe that the proposed method offers the potential foundation for terahertz non-fluorescent microscopy and sample detection with weak scattering.

**Funding**. National Key Research and Development Program of China (2019YFA0210203); National Natural Science Foundation of China (61971013, 61531002).